# From Physical Difference to Meaning: A Constructor-Theoretic Framework for Prebiotic Information in Casimir-Lifshitz-Coupled Protocell Clusters

Michael Massoth
Department of Computer Science, Hochschule Darmstadt (h_da)
University of Applied Sciences Darmstadt, member of European University of Technology (EUt+)
Darmstadt, Germany
e-mail: michael.massoth@h-da.de

*Abstract*- **This paper develops a physical framework for the prebiotic emergence of information and meaning. Building on Constructor Theory, we define information as a reproducible physical difference and meaning as a difference with stable functional consequences. Casimir–Lifshitz–coupled protocell clusters serve as a minimal model that exhibits reproducible attractors, ordered transitions, and autonomous task structures. We show that such clusters carry both informational states (e.g., distances, geometries, gradients) and meaningful states that regulate prebiotic tasks such as approach, exchange, or stabilization. This approach integrates physical mechanisms, computational mechanics, and early proto-semantic functions into a coherent account of information formation before biology.**

I. INTRODUCTION

This is the third of seven papers in the series: "A Constructivist Proto-Bio-Information Theory: A Physically Grounded Nano-Systems Architecture for Prebiotic Emergence, Information, Proto-Semantic Function, and Sustainability of Protocell Aggregation and Cluster Formation".

Massoth [1] shows that Casimir–Lifshitz forces generate robust attraction and stable protocell clusters across 5–200 nm under realistic prebiotic conditions. These physically enforced mesoscale assemblies form the structural substrate in which this paper grounds the emergence of information and meaning.

Massoth [2] demonstrates that such clusters form reproducible mesoscale attractors with autonomous ε-machine dynamics. This attractor-based organization provides the computational and informational foundation that this paper develops into a physical account of proto-information and proto-semantics.

This work applies the Constructor Theory of Information proposed by Deutsch and Marletto [4] to a computational-mechanics framework of ε-machines and attractor-based memory developed by Rosas *et al.* [3], establishing a unified physical and informational architecture for protocell-cluster dynamics.

The structure of the paper is as follows: In Section II, we introduce the motivation, conceptual gap, and guiding research questions, framing prebiotic information and meaning within a constructor-theoretic perspective. Section III reviews related work on semantic information, protocells, and Constructor Theory, positioning our approach relative to existing biological and computational models.

Section IV outlines the core principles of Constructor Theory, with a focus on tasks, constructors, and the physical definition of information. In Section V, we formalize information as a reproducible physical difference and demonstrate how protocell cluster attractors satisfy constructor-theoretic information criteria.

Section VI develops the central concept of meaning as a functionally exploited difference, showing how specific protocell states control downstream task realizations. Section VII applies this framework to Casimir–Lifshitz–coupled protocell clusters, identifying concrete tasks and constructor-like behaviors.

Finally, Section VIII summarizes the conceptual contributions, answers the guiding research questions, relates the framework to Computational Mechanics, and discusses implications, limitations, and future research directions.

II. MOTIVATION, RESEARCH QUESTION AND RELEVANCE

*A. Motivation: Information and Meaning in Prebiotic Systems*

How prebiotic systems generated information, function, and eventually meaning before genes, enzymes, or metabolism emerged remains a central challenge in origins-of-life research. Classical models describe self-organization yet offer little insight into when a physical pattern qualifies as information or under which conditions differences become functionally consequential. The distinction between information formation and meaning typically appears only once biological codes exist.

Here we propose an alternative approach. Constructor Theory defines information through physically possible tasks: information is a reproducible physical difference. Meaning arises when such a difference produces stable functional consequences that influence processes such as persistence or reproduction. Prebiotic information thus becomes a physical question: Which tasks were allowed by early Earth conditions, which substrates could support them, and when did differences begin to be used rather than merely produced?

*B. Approach, Gap, and Guiding Research Questions*

We apply Constructor Theory to prebiotic protocells that interact through Casimir–Lifshitz forces and electrolyte-mediated fields. In contrast to the accompanying Computational Mechanics paper, which analyzes ε-machines, we focus here on meaning and function. Our framework proceeds in three steps: core principles of Constructor Theory; information as a reproducible physical difference; and meaning as a difference functionally exploited within prebiotic task networks.

Current models of chemical self-organization rarely specify which informational tasks were physically feasible, when differences are merely structural patterns, or when they become functionally active. It also remains unclear at what point protocells qualify as constructors capable of executing repeatable tasks.

This motivates three guiding research questions:

*RQ#1: Under which physical conditions can protocell clusters act as constructors that repeatedly perform tasks such as approach, exchange, or coupling?*

*RQ#2: How can pure informational differences be distinguished from meaningful states with distinct functional outcomes?*

*RQ#3: Which classes of tasks in Casimir-coupled clusters can support proto-semantic stability, and how does this relate to concepts such as informational and causal closure?*

*C. Relevance for Biology and Computer Science*

Our approach is deliberately interdisciplinary. In biology, protocells appear not only as reaction compartments but as early functional units capable of performing tasks without genes or enzymes. In computer science, Constructor Theory broadens the notion of a program to networks of physical tasks, offering new perspectives on natural information processing.

## III. RELATED WORK

Theories of semantic information define meaning via viability or functional relevance, yet often operate at the level of abstract agents or single synthetic cells. Experimental work such as Ruzzante et al., "Synthetic Cells Extract Semantic Information From Their Environment" (2023) [24] demonstrates semantic information processing, but does not address its prebiotic physical realization [23] [25]. Constructor Theory of Information from D. Deutsch and C. Marletto, "Constructor Theory of Information" [4] provides a physical definition of information as a reproducible difference, but has rarely been applied to origins-of-life systems.

This paper integrates these strands by applying Constructor Theory to Casimir–Lifshitz–coupled protocell clusters whose ε-machine organization was established in [2]. Information is defined as a reproducible physical difference; meaning arises when such differences systematically enable or disable tasks. Protocell clusters thus act as partial constructors, in which cluster states regulate prebiotic tasks such as approach, exchange, or stabilization. Meaning is thereby grounded in physical task structure rather than symbolic representation.

## IV. CORE PRINCIPLES OF CONSTRUCTOR THEORY

Constructor Theory (CT) of information [4] describes physical processes in terms of tasks—transformations that are possible, impossible, or only conditionally realizable. Instead of trajectories, CT focuses on transitions that can, in principle, be reproduced indefinitely. This makes CT well suited for prebiotic protocells, where information and early meaning arise not through symbols but through physically achievable transformations.

*A. Tasks and Constructors*

A task $A=\{x_i \rightarrow y_i\}$ maps input attributes of a substrate to output attributes. These attributes may be geometric, chemical, or energetic, such as distances, field configurations, or cluster shapes. A constructor is a system that performs a task repeatedly and reliably without losing its ability to do so. Perfect stability is unnecessary; what matters is the theoretical possibility of arbitrarily increasing accuracy. Enzymes, technical protocols, and stable protocell clusters all function as constructors because they transform states reproducibly without being consumed. CT thus generalizes concepts across chemistry, biology, and computer science. In a prebiotic context, protocells can be viewed as natural constructors that translate environmental differences into stable internal or collective states.

*B. Possible versus Impossible Tasks*

CT expresses physical laws as constraints on which tasks are permitted. A task is possible if physics allows a constructor to perform it with arbitrarily high accuracy. It is impossible if it would violate fundamental principles such as energy conservation or the quantum no-cloning theorem. CT therefore studies the space of what can occur, not detailed time evolutions. This is valuable for protocell clusters: their microscopic dynamics are complex, but mesoscale processes—approach, exchange, stabilization—form well-defined task classes.

Casimir–Lifshitz–coupled protocells thus support a set of physically allowed tasks arising from interaction forces, geometry, and thermal fluctuations. CT provides a unified vocabulary without requiring a full microphysical derivation.

*C. Constructor Theory of Information*

In CT, a substrate carries information when its attributes are distinguishable and copiable through physically possible tasks. Distinguishability requires a task that maps two attributes to different, readable outputs while the constructor remains functional. Copyability requires a task that transfers a state onto a second, initially blank substrate.

Many physical systems—such as general quantum states—do not satisfy this copying requirement. Information in CT is therefore a reproducible physical difference, not a symbolic one. Protocell clusters exhibit such differences through stable distances, geometries, and field states. A classical bit is defined not by symbols "0/1" but by the existence of a task family capable of generating, copying, and transforming those differences.

## V. INFORMATION AS A REPRODUCIBLE PHYSICAL DIFFERENCE

### A. From Symbolic to Physical Notions of Information

The Shannon framework describes information in terms of probabilities, encoding, and channel transmission, assuming senders, receivers, and symbolic structures—features absent in prebiotic environments. Constructor Theory replaces this symbolic view with a physical one: information is a reproducible difference in the world, independent of interpretation or statistics.

A difference between two states counts as information only if physically possible tasks exist that can generate, copy, transmit, or read it with arbitrarily increasing accuracy. Information is therefore a property of substrates and their space of possible transformations, not merely of descriptions. The key question shifts from "How many bits?" to "Which differences can be reliably realized in this world?"

This perspective aligns with results from [2]: mesoscale attractors in protocell clusters form reproducible macro-states with stable, recurrent transitions, robust to microscopic noise [10]. These attractors satisfy the CT criteria for informational states.

### B. Reproducible Differences as Tasks

In Constructor Theory, reproducible differences can be described directly as tasks. The most elementary task is *distinguishing*: $(x_A \rightarrow y_A), (x_B \rightarrow y_B)$ with $y_A \neq y_B$.

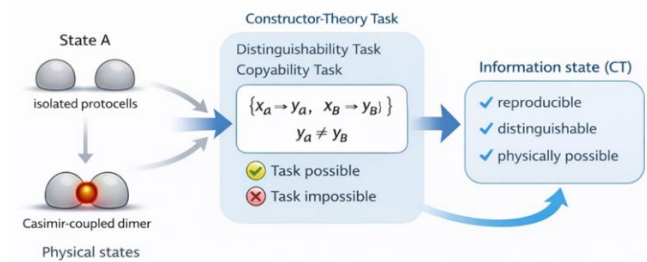

Figure 1. From physical difference to information in Constructor Theory.

In Figure 1, two physical protocell states are shown, isolated vesicles and a Casimir-coupled dimer. Only when physically possible distinguishability and copyability tasks exist does a mere physical difference qualify as an information state in Constructor Theory, emphasizing reproducibility, distinguishability, and physical realizability.

If such a task is physically possible, then $x_A$ and $x_B$ are distinguishable, and the difference $\Delta(x_A, x_B)$ carries information. A difference counts as information only if the laws of nature permit its reproducible realization. Random, unstable, or purely formal differences do not constitute information unless they can be reliably instantiated [13].

The ε-machine analysis from Paper 2 supports this criterion. Protocell clusters satisfy *informational closure*: macrostates of cluster configurations predict future states better than the full microscopic dynamics. Differences between attractors are therefore not only reproducible but also causally effective - strong evidence for information in the CT sense [11].

### C. Protocellular Examples of Reproducible Differences

Protocell cluster systems present several classes of prebiotic differences that satisfy CT information criteria and are further supported by results from [2][15].

*Isolated vs. Coupled Protocells:*
The transition from isolated protocells to dimers is robustly generated by Casimir–Lifshitz forces, reproducible, and forms a well-defined macro-state. The associated distance signature (e.g., L ≈ 5–20 nm) can be read out through physically possible tasks, such as changes in local field structures or membrane deformations. In the ε-machine model, this difference corresponds to distinct causal states $E_i$, and the transition follows defined paths $T_{ij}$ - precisely the structure expected for a CT informational difference.

*(2) Distinct Cluster Geometries (tetrahedral, octahedral, icosahedral):*
These attractors differ clearly in contact graphs, symmetry, energetic minima, and return dynamics. Identified as stable attractors in Paper 2, they meet CT requirements for distinguishable and reproducibly realizable states.

The attractor-based memory- formally $\|Z_{t+1} - A\| < \|Z_t - A\|$, - provides the stability CT demands for reproducible differences.

*(3) Ion gradients and electrochemical differences:*
Protocells capable of maintaining consistently high versus low ion gradients realize a difference that can be copied (via gradient transport) or distinguished (via membrane responses). Such differences define an informational variable, provided that $(x_{high} \rightarrow y_{high}), (x_{low} \rightarrow y_{low})$ are physically possible tasks.

*(4) ε-machine states as informational differences:*
A major result from [2] is that macro-states $Z_t$ differ not only geometrically but through their future-state distributions. Two states belong to different ε-state classes $E_i$ and $E_j$ when $P(Z_{t+1} | Z_t \in E_i) \neq P(Z_{t+1} | Z_t \in E_j)$.

From a CT perspective, this implies that tasks exist that can not only generate these differences but also exploit them functionally, because different transitions lead to different consequences.

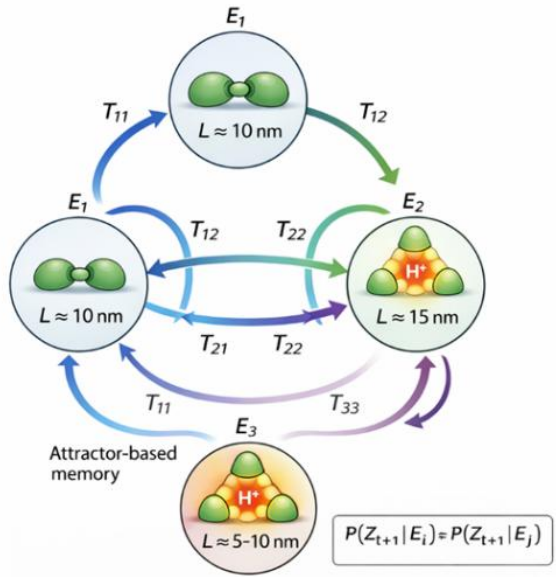

Figure 2. ε-machine attractors with geometry-dependent proton-gradient states.

In Figure 2, an ε-machine network illustrates how distinct protocell-cluster attractors encode reproducible informational states. Dimers and tetrahedral clusters form macro-states with characteristic Casimir separations L and transition pathways. Central proton ($H^+$) gradients in tetrahedral states indicate geometry-dependent quantum-field-induced energy reservoirs underlying predictive state differentiation.

*Interim Conclusion:* A difference qualifies as information in Constructor Theory only when it is physically reproducible. Results from [2] show that protocell clusters possess precisely such reproducible differences: stable attractors, robust transitions, attractor-based memory, and structured ε-machines. They therefore satisfy the necessary criteria to serve as prebiotic information carriers in the CT framework.

## VI. MEANING AS A FUNCTIONALLY EXPLOITED DIFFERENCE

### A. From Information to Meaning: Additional Conditions

In Constructor Theory, information is a reproducible physical difference [4][16] that can be generated, copied, or read through tasks. This alone does not constitute meaning. The crucial addition is functional: meaning is information that plays a stable role within a network of tasks.
An informational difference ($x_A$, $x_B$) gains meaning when it systematically determines which tasks a system performs and what consequences follow [23].
The substrate then acts not only as a variable but as a controller of possible transformations. A difference has meaning when constructors exist whose tasks are realizable only for specific informational values:
$x_A \Rightarrow T_A$, $x_B \Rightarrow T_B$, $T_A \neq T_B$.

In this sense, information ($x_A$ or $x_B$) acts as a control variable. Paper [2] provides empirical support: the ε-machines of prebiotic protocell clusters exhibit distinct causal states with different transition patterns and functional outcomes. This mapping from state to transformation is exactly how CT defines meaning.

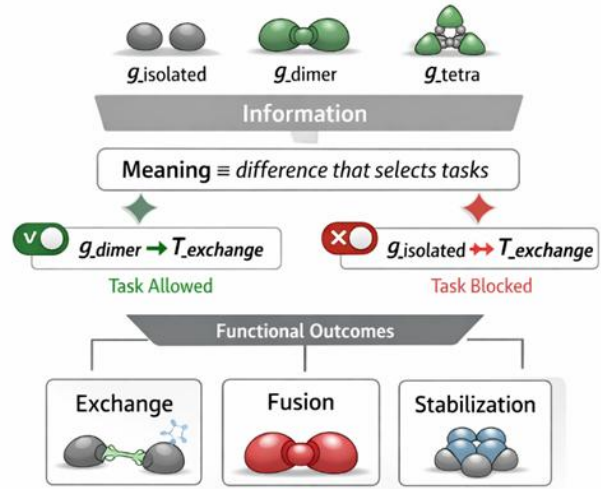

Figure 3. From information to meaning via task-gating in protocell clusters

Figure 3, Schematic illustrating the transition from information to meaning via task-gating in Casimir–Lifshitz–coupled protocell clusters. Cluster states act as informational variables, while meaning emerges when specific states selectively enable or block downstream tasks. Meaning is thus defined as a physical difference that controls functional transformations.

### B. Task Networks and Functional Roles

Physical systems consist of chained task networks, where the output of one task becomes the input of another. Within such networks, functional roles emerge: some states act as triggers initiating tasks; others act as switches selecting pathways; still others serve as signals guiding subsystem responses.
Formally, a task is possible only if the relevant informational variable has the appropriate value. A state gains meaning when it enables or restricts the realizability of other tasks.
For biologists, a meaningful state modulates adaptive behavior or responses. For computer scientists, meaning parallels the role of a bit in program flow, such as a conditional branch. For sustainability science, meaning resembles the role of a state within a regulatory system. Massoth [2] offers a prebiotic realization of these roles: the ε-machine of the cluster specifies which transitions are possible and thus which functional consequences arise.

### C. Prebiotic Emergence of Meaning in Protocell Clusters

The central question is whether simple Casimir–Lifshitz–coupled protocells can already realize functional roles and thereby generate proto-semantic structures [12]. The answer is yes, provided that differences in cluster states systematically enable different tasks.

*(1) Coupled vs. Isolated Protocells as a Functional Difference:*

The distinction between isolated protocells and dimers (or higher configurations such as tetrahedra or octahedra) is established in Paper 2 as a stable informational state with geometry variable G={$g_{isolated}$,$g_{dimer}$,$g_{tetra}$,...}. This difference acquires meaning when it modulates downstream tasks—for example:

(a) increased likelihood of fusion,
(b) enhanced molecular exchange due to reduced separation,
(c) altered collective membrane modes or field configurations.

If a task such as an exchange process $T_{exchange}$ is possible only in the dimer state, Dimer ⇒ $T_{exchange}$, isolated ⇏ $T_{exchange}$, then the state "dimer'' carries meaning. This is precisely the constructor-theoretic condition.

The ε-machine analysis in Paper 2 confirms this: dimer states form distinct causal classes $E_i$ and evolve differently from isolated states, giving rise to different functional consequences—i.e., meaning.

*(2) Casimir-shaped Distances and Field Configurations as Meaningful Differences:*

Casimir–Lifshitz forces generate discrete, energetically preferred separations. Distances and orientations thus act as informational variables, such as distance variable L={$L_{near}$, $L_{far}$} or gradient variable D={$\Delta_{low}$,$\Delta_{high}$}. These differences acquire meaning when they enable different tasks, for example:

• higher reaction rates at $L_{near}$,
• more efficient energy transfer through collective field modes,
• stronger synchronization of ionic gradients.

If the set of possible tasks depends on the realized distance- $L_{near}$⇒ $T_{fusion}$, $L_{far}$⇒ $T_{diffusion}$,
then the distance difference is functional, and therefore meaningful.

Massoth [2] shows that these separations are attractor states stabilized by attractor-based memory and that they induce distinct transitions $T_{ij}$. These differentiated paths constitute the functional consequences that define meaning in Constructor Theory (CT).

*Interim Conclusion:* Meaning arises when physical differences are not only reproducible but also functionally effective. Casimir-stabilized protocell clusters satisfy both conditions. They carry informational states through stable attractors and ε-machine–structured transitions, and meaningful states through systematic modulation of prebiotic tasks such as approach, exchange, fusion, stabilization, or gradient formation. Such clusters therefore represent a minimal physical context in which proto-semantic structures can emerge.

VII. APPLICATION: CASIMIR–LIFSHITZ–COUPLED PROTOCELL-CLUSTERS AS CONSTRUCTORS

We now apply the constructor-theoretic framework of information to prebiotic protocell systems that form mesoscale clusters through Casimir–Lifshitz attraction and electrolyte-mediated coupling. Paper [2] shows that these clusters implement an ε-machine with stable attractors and structured transition pathways, generating an autonomous macro-level with attractor-coded memory. These properties make protocell clusters natural candidates for constructor-theoretic tasks.

*A. Casimir–Lifshitz–Coupled Protocell-Clusters as Physical Substrates*

Protocells with radii of $R \approx$ 100–1000 nm interact in saline prebiotic environments via Casimir–Lifshitz forces. The resulting potential landscapes contain stable and metastable minima, giving rise to long-lived configurations such as dimers, tetrahedral clusters, or 13-membered icosahedra. Massoth [2] demonstrates that these clusters form robust attractors with characteristic binding energies and structured transition dynamics. They satisfy informational closure, as future states are better predicted from macro-states than from full microscopic data.

In Constructor Theory, this implies that protocell clusters provide physical substrates carrying stable attributes—distances, geometries, contact graphs, and gradients—and reproducible transformation pathways. These attributes and transitions supply the material basis for possible tasks.

*B. Elementary Tasks in Protocell Clusters*

Under prebiotic conditions, three fundamental tasks can be identified that Casimir-coupled clusters can perform repeatedly and robustly [14]. The ε-machine analysis from Paper 2 confirms their structural stability.

*Task A: Approach and binding of two protocells:*
($x_{isolated} \to x_{dimer}$) with A={$g_{isolated} \to g_{dimer}$}
Two vesicles with suitable orientation, located in a region where the Casimir–Lifshitz force is net attractive, reliably relax into a stable dimer separation *L* (surface-to-surface distance). The ε-machine encodes this transition as a strictly causal edge $E_{isolated} \to E_{dimer}$.

*CT interpretation:* If this transition can be realized with arbitrarily increasing accuracy in principle, Task A is possible (A✓).

*Task B: Exchange and modification of gradients within the dimer:* ($x_{gradient} \to x_{gradient'}$) with C={($\Delta_{low} \to \Delta'_{high}$)}
Once a dimer is formed, contact regions and field couplings enable repeated exchange of ions or molecules. Massoth [2] shows that dimers persist long enough to develop attractor-based memory, making these exchange patterns reproducible.

*CT interpretation:* The dimer functions as a constructor that transforms gradients.

*Task C: Stabilization of specific cluster configurations:*

($X \to X$) over $\Delta t$

A cluster maintains its geometry despite thermal fluctuations by relaxing back to its attractor along characteristic return trajectories.

*CT interpretation:* This structural persistence constitutes a maintenance task enabled by autonomous attractor memory [2], defining a maintenance constructor.

### C. Protocell Clusters as (Partial) Constructors

A system qualifies as a constructor if it can perform a task repeatedly without losing its ability to do so. Protocell clusters meet this criterion. Casimir–Lifshitz attraction stabilizes repeated approach events; long lifetimes support robust exchange processes; and the ε-machine defines reproducible transitions and return pathways [24]. The cluster thus transforms input states into output states while retaining its capacity for task execution.

Protocells are not intentional agents; their constructor capacity emerges from physical law. They may be considered partial constructors, since some resources—such as temperature or ion supply—derive from the environment. Constructor Theory permits such side effects as long as the constructor determines the structure of the transformation.

### D. Information vs. Meaning States in Protocell Clusters

Constructor Theory provides a natural classification of the cluster states identified in [2]. Information states are physical differences for which tasks of generation, copying, or discrimination exist. These include monomers, dimers, and higher-order clusters, distinct ion gradients, defined distance and field configurations, and ε-machine attractors. Information states satisfy: $(x_A \neq x_B) \wedge A\checkmark, B\checkmark$.

*Meaning states* are those information states that trigger different downstream tasks. A dimer, for example, can enable further approach or gradient modification. Smaller separations can activate transport or fusion tasks that are impossible at larger separations. Distinct attractors lead to different transition pathways. A meaning state satisfies the constructor-theoretic condition that different information values realize systematically different tasks: $x_A \Rightarrow T_i$, $x_B \Rightarrow T_j$, $T_i \neq T_j$.

*Interim conclusion:* Casimir–Lifshitz–coupled protocell clusters are both physical aggregates and natural constructors that implement reproducible tasks such as approach, exchange, and stabilization. They carry information states as reproducible differences and meaning states as differences with functional consequences in the task network. They thus realize the constructor-theoretic transition from information to meaning in a concrete prebiotic setting and complement the mechanisms of informational closure and attractor-based memory described in [2].

## VIII. CONCLUSION AND FUTURE WORK

### A. Summary of the Conceptual Contribution

This work links Constructor Theory to a realistic prebiotic scenario. In this framework, information is a reproducible physical difference, and meaning arises when such a difference has stable functional consequences within a network of tasks. Casimir–Lifshitz–coupled protocell clusters serve as a minimal model: their stable attractors form reproducible information states, and their transition and return pathways enable basic tasks such as approach, exchange, and stabilization. Physical differences thus become functional categories.

Our contribution is twofold: information is the capacity of a substrate to realize specific tasks and meaning emerges when these differences systematically guide transformations.

### B. Answering the research questions

*RQ#1: Under which physical conditions can protocell clusters act as constructors that repeatedly perform tasks such as approach, exchange, or coupling?*

Protocell clusters function as repeatable constructors when nanoscale separations, dielectric contrasts and moderate ionic strengths jointly create Casimir–Lifshitz–dominated interaction landscapes with discrete, thermally stable attractors. Under these conditions, vesicles can repeatedly approach, bind and reconfigure without structural degradation, and the resulting attractor network provides an autonomous mesoscale level capable of supporting reproducible tasks such as coupling, exchange and geometric stabilization.

*RQ#2: How can pure informational differences be distinguished from meaningful states with distinct functional outcomes?*

Pure informational differences correspond to reproducible physical distinctions—cluster geometries, separation states or ionic gradients—that can, in principle, be generated, discriminated or copied. These differences become meaningful when they systematically change which downstream tasks are possible and thus produce distinct functional consequences. In Casimir-coupled protocells, meaning arises when particular spatial or gradient states act as control variables that gate transformations such as fusion, exchange or attractor maintenance.

*RQ#3: Which classes of tasks in Casimir-coupled clusters can support proto-semantic stability, and how does this relate to concepts such as informational and causal closure?*

Proto-semantic stability is supported by three mutually reinforcing task classes: approach and coupling tasks that establish persistent relational structures; exchange and transformation tasks that reorganize internal gradients in a state-dependent manner; and maintenance tasks that restore perturbed configurations to their characteristic attractors. Together, these tasks form a mesoscale dynamic that is informationally and causally closed, meaning that future transitions are best predicted from macro-states rather than microstates. Within this closed task network, certain informational states reliably select—and are stabilized by—specific transformations, constituting a physically grounded precursor to semantic organization in prebiotic protocell systems.

*C. Relation to Computational Mechanics*

This approach complements paper [2], which describes protocell clusters as ε-machines whose causal states encode attractors and their mesoscale switching pathways. This macro-level is informationally and causally closed and displays a proto-software-like dynamic.

Here, we ask which of these states carry information and which acquire meaning. Computational Mechanics characterizes structure and predictability; Constructor Theory characterizes tasks and functional roles. Together, they show how ε-machines identify stable differences while CT determines which become functionally effective.

*D. Relevance for Biology and Computer Science*

The framework offers a prebiotic definition of function independent of genes or enzymes. Protocells appear as active physical units that stabilize distances, generate gradients, and modulate exchange processes. Function arises when such differences reliably trigger or regulate tasks.

In computer science, the approach broadens the notion of a program: programs become networks of physical tasks guided by state variables. Protocell clusters thus implement a natural, substrate-bound computation, where information states act as switches and meaning states as control structures.

*E. Limitations and Future Work*

The contribution is conceptual. The proposed information and meaning states require experimental and numerical validation. Models of Casimir–Lifshitz potential landscapes and mesoscale dynamics could test the robustness of approach, exchange, and stabilization tasks under realistic conditions. Experiments with protocell models—vesicles or coacervates—may probe analogous attractors and their functional consequences.

A future integrated theory should unify Computational Mechanics and Constructor Theory to determine which macrostates function as constructors and how meaning states emerge from ε-machine organization. This work represents a first step toward a physical explanation of how stabilized differences gave rise to functional categories and early proto-semantic structures.